\documentstyle[12pt]{article}
\catcode`\@=11

\@addtoreset{equation}{section}
\def\theequation{\arabic{section}.\arabic{equation}}

\catcode`\@=11

\def\section{\@startsection{section}{1}{\z@}{3.5ex plus 1ex minus
   .2ex}{2.3ex plus .2ex}{\large\bf}}

\def\eqnarray{\let\@currentlabel=\theequation\refstepcounter{equation}
    \global\@eqnswtrue
    \global\@eqcnt\z@\tabskip\@centering\let\\=\@eqncr
    $$\halign to \displaywidth\bgroup\@eqnsel\hskip\@centering
      $\displaystyle\tabskip\z@{##}$&\global\@eqcnt\@ne
       \hfil${{}##{}}$\hfil
      &\global\@eqcnt\tw@ $\displaystyle\tabskip\z@{##}$\hfil
       \tabskip\@centering&\llap{##}\tabskip\z@\cr}
\def\lefteqn#1{\hbox to 4\arraycolsep{$\displaystyle #1$\hss}}
\def\thesection{\arabic{section}.}

\def\appendix{\setcounter{section}{0}
        \def\thesection{Appendix.}
        \def\theequation{\Alph{section}.\arabic{equation}}}

\long\def\@makefntext#1{\parindent 0cm\noindent
\hbox to 1em{\hss$^{\@thefnmark}$}#1}
\def\rref#1{(\ref{#1})}
\newcommand{\beq}{\begin{equation}}
\newcommand{\eeq}{\end{equation}}
\begin{document}
%%%%%%%%%%%%%%%%%%%%%%%%%%%%%%%%%%%%%%%%%%%%%%%%%%%%%%%%%%%%%%%%%%%%%%%%%%%
%     C I T E . S T Y
%     compressed lists of numerical citations: [11-16]
%     see also OVERCITE.STY and DRFTCITE.STY
%
%     Copyright (C) 1989-1992 by Donald Arseneau
%     These macros may be freely transmitted, reproduced, or modified for
%     non-commercial purposes provided that this notice is left intact.
%
%
%  \@citen contains the code that parses the list of names, ignoring
%  spaces after commas, writes the aux file \citation, and formats the
%  number list.  \citen can be used by itself to give citation numbers
%  without the other formatting; e.g., "See also ref.~\citen{junk}."
%
\def\citen#1{%
\edef\@tempa{\@ignspaftercomma,#1, \@end, }% ignore spaces in parameter list
\edef\@tempa{\expandafter\@ignendcommas\@tempa\@end}%
\if@filesw \immediate \write \@auxout {\string \citation {\@tempa}}\fi
\@tempcntb\m@ne \let\@h@ld\relax \let\@citea\@empty
\@for \@citeb:=\@tempa\do {\@cmpresscites}%
\@h@ld}
%
% for ignoring spaces in the input:
\def\@ignspaftercomma#1, {\ifx\@end#1\@empty\else
   #1,\expandafter\@ignspaftercomma\fi}
\def\@ignendcommas,#1,\@end{#1}
%
% For each citation, check if it is defined, if it is a number, and
% if it is a consecutive number that can be represented like 3-7.
%
\def\@cmpresscites{%
 \expandafter\let \expandafter\@B@citeB \csname b@\@citeb \endcsname
 \ifx\@B@citeB\relax % undefined
    \@h@ld\@citea\@tempcntb\m@ne{\bf ?}%
    \@warning {Citation `\@citeb ' on page \thepage \space undefined}%
 \else%  defined
    \@tempcnta\@tempcntb \advance\@tempcnta\@ne
    \setbox\z@\hbox\bgroup % check if citation is a number:
    \ifnum\z@<0\@B@citeB \relax
       \egroup \@tempcntb\@B@citeB \relax
       \else \egroup \@tempcntb\m@ne \fi
    \ifnum\@tempcnta=\@tempcntb % Number follows previous--hold on to it
       \ifx\@h@ld\relax % first pair of successives
          \edef \@h@ld{\@citea\@B@citeB}%
       \else % compressible list of successives
%         % use \hbox to avoid easy \exhyphenpenalty breaks
          \edef\@h@ld{\hbox{--}\penalty\@highpenalty \@B@citeB}%
       \fi
    \else   %  non-successor--dump what's held and do this one
       \@h@ld \@citea \@B@citeB \let\@h@ld\relax
 \fi\fi%
 \let\@citea\@citepunct
}
%
%%    To put space after the comma, use:
\def\@citepunct{,\penalty\@highpenalty\hskip.13em plus.1em minus.1em}%
%%    For no space after comma, use:
%% \def\@citepunct{,\penalty\@highpenalty}%
%%
%
%  Make \@citex refer to \citen:
%
\def\@citex[#1]#2{\@cite{\citen{#2}}{#1}}%
%
%  Replacement for \@cite.  Give one normal space before the citation,
%  set high penalties for linebreaks,
%
\def\@cite#1#2{\leavevmode\unskip
  \ifnum\lastpenalty=\z@ \penalty\@highpenalty \fi % highpenalty before
  \ [{\multiply\@highpenalty 3 #1% % triple-highpenalties within list
      \if@tempswa,\penalty\@highpenalty\ #2\fi % and before note.
    }]\spacefactor\@m}
\let\nocitecount\relax  % in case \nocitecount was used for drftcite
%
%%%%%%%%%%%%%%%%%%%%%%%%%%%%%%%%%%%%%%%%%%%%%%%%%%%%%%%%%%%%%%%%%%%%%%%%%%
\begin{titlepage}
\vspace{.5in}
\begin{flushright}
UCD-96-13\\
June 1996\\
revised September 1996\\
gr-qc/9606043\\
\end{flushright}
\vspace{.5in}
\begin{center}
{\Large\bf
 The Statistical Mechanics of the\\[1.2ex]
 Three-Dimensional Euclidean Black Hole}\\
\vspace{.4in}
{S.~C{\sc arlip}\footnote{\it email: carlip@dirac.ucdavis.edu}\\
       {\small\it Department of Physics}\\
       {\small\it University of California}\\
       {\small\it Davis, CA 95616}\\{\small\it USA}}
\end{center}

\vspace{.5in}
\begin{center}
\begin{minipage}{4in}
\begin{center}
{\large\bf Abstract}
\end{center}
{\small\hspace*{.8em}
In its formulation as a Chern-Simons theory, three-dimensional general
relativity induces a Wess-Zumino-Witten action on spatial boundaries.
Treating the horizon of the three-dimensional Euclidean black hole as
a boundary, I count the states of the resulting WZW model, and show
that when analytically continued back to Lorentzian signature, they
yield the correct Bekenstein-Hawking entropy.  The relevant states can
be understood as ``would-be gauge'' degrees of freedom that become
dynamical at the horizon.}
\end{minipage}
\end{center}
\end{titlepage}
\addtocounter{footnote}{-1}

The underlying microscopic source of black hole entropy is not yet
understood, but it is natural to conjecture that it originates in
quantum gravitational degrees of freedom at the black hole horizon.
If this is true, however, then the (2+1)-dimensional black hole of
Ba{\~n}ados, Teitelboim, and Zanelli \cite{BTZ} presents a paradox.
General relativity in three spacetime dimensions can be rewritten as
a Chern-Simons theory \cite{Achu,Witten}, and as such, its degrees
of freedom are fairly well understood.  In particular, the small
number of topological degrees of freedom of a Chern-Simons theory
are not sufficient to account for the large entropy of a macroscopic
BTZ black hole.

In reference \cite{Car:stat}, and independently in \cite{Bal}, a
possible solution to this paradox was suggested.  A Chern-Simons
theory on a manifold with boundary induces a Wess-Zumino-Witten
theory on the boundary \cite{EMSS,Wita}, and this WZW model can have
many more degrees of freedom than the original Chern-Simons theory.
These new degrees of freedom are ``would-be pure gauge'' excitations
that become physical at the boundary \cite{Ogura,Car:Liou}.  Let us
suppose that the event horizon of a black hole can be treated as a
boundary.  (I will return to this assumption later.)  Then the induced
WZW model at the horizon offers a natural source of microscopic degrees
of freedom.

A preliminary counting argument in reference \cite{Car:stat} indicated
that these degrees of freedom can correctly account for the entropy of
the BTZ black hole.  That analysis was based on plausible but unproven
assumptions about the quantization of the $\hbox{SU}(1,1)$ WZW
model.  In this paper, I consider the analytic continuation to the
better-understood $\hbox{SL}(2,{\bf C})$ WZW model obtained from Euclidean
gravity in three dimensions, and demonstrate that the entropy of the
three-dimensional black hole can be derived as the logarithm of the number
of microscopic states at the horizon. The results are quite robust: the
semiclassical contribution to the entropy is determined by Virasoro
zero-modes, and is independent of the details of the rest of the Hilbert
space.

\section{``Would-be Gauge'' Degrees of Freedom}

Before proceeding with the computation, it is useful to recall the
source of boundary degrees of freedom in Chern-Simons theory \cite{PCGC}.
Consider a Chern-Simons theory on a manifold with boundary, described by
the action
\beq
I_{\hbox{\scriptsize CS}} = {k\over4\pi}\int_M
  \hbox{Tr}\left( A\wedge dA + {2\over3}A\wedge A\wedge A \right)
  + {k\over4\pi}\int_{\partial M} \hbox{Tr}\,A_zA_{\bar z} .
\label{b1}
\eeq
The boundary term in \rref{b1} is the one appropriate for fixing the
field $A_z$ at $\partial M$.  If $M$ is closed, this term disappears,
and $e^{i I_{\hbox{\tiny CS}}}$ is gauge invariant. If $M$ has a boundary,
however, this invariance is broken.  Indeed, under the decomposition
\beq
A = g^{-1}dg + g^{-1}\tilde A g ,
\label{b2}
\eeq
the action becomes \cite{Ogura,Car:Liou}
\beq
I_{\hbox{\scriptsize CS}}[A]
   = I_{\hbox{\scriptsize CS}}[\tilde A]
   + k I^+_{\hbox{\scriptsize WZW}}[g,\tilde A_z] ,
\label{b3}
\eeq
where $I^+_{\hbox{\scriptsize WZW}}[g,\tilde A_z]$ is the action of
a chiral WZW model on the boundary $\partial M$,
\beq
I^+_{\hbox{\scriptsize WZW}}[g,\tilde A_z]
 = {1\over4\pi}\int_{\partial M}\hbox{Tr}
 \left(g^{-1}\partial_z g\,g^{-1}\partial_{\bar z} g
 - 2g^{-1}\partial_{\bar z} g {\tilde A}_z\right)
 + {1\over12\pi}\int_M\hbox{Tr}\left(g^{-1}dg\right)^3 .
\label{b4}
\eeq
The ``pure gauge'' degrees of freedom $g$ are thus promoted to true
dynamical degrees of freedom at the boundary.

A similar phenomenon can occur in general relativity.  The infinitesimal
analog of the decomposition \rref{b2} may be obtained by performing a
transverse splitting of small fluctuations of a background metric
$g_{\mu\nu}$,
\beq
\delta g_{\mu\nu} = h_{\mu\nu} + (K\xi)_{\mu\nu} ,\quad
  (K^\dagger h)_\mu = 0 ,\qquad\hbox{with}\quad 
(K\xi)_{\mu\nu} = \nabla_\mu\xi_\nu + \nabla_\nu\xi_\mu
  = {\cal L}_\xi g_{\mu\nu} .
\label{b6}
\eeq
If $M$ is closed, this splitting is unique \cite{Ebin,York}, and
provides a standard division into ``physical'' and ``gauge'' degrees
of freedom.  If $M$ has a boundary, however, a unique decomposition
requires boundary conditions that make $K^\dagger K$ self-adjoint.
The simplest choice is
\beq
\left. \xi^{\mu}\right|_{\partial M} = 0 .
\label{b7}
\eeq
Once again, the ``would-be gauge'' degrees of freedom $K\xi$ with
$\xi^\mu\ne 0$ at $\partial M$ are potential new dynamical degrees of
freedom at the boundary.  Evidence for this special role comes from
the canonical formalism \cite{Bal}: the generator of the transformation
$g_{ij}\rightarrow g_{ij} + (K\xi)_{ij}$ on a spacelike hypersurface
$\Sigma$ is proportional to a constraint, and thus generates a symmetry,
only when $\xi^i$ vanishes at $\partial\Sigma$.
Note, of course, that vector fields $\xi$ in the kernel of $K$, the
Killing vectors of $g_{\mu\nu}$, do {\em not\/} give rise to new degrees
of freedom.  This will be important to the later analysis.

Unfortunately, the decomposition \rref{b6} holds only for infinitesimal
variations $\delta g_{\mu\nu}$; the finite version is highly nonlocal.
The gravitational analog of the WZW action is consequently difficult
to find (although see \cite{Teit}).  In three spacetime dimensions,
however, we can avoid this difficulty: three-dimensional general
relativity can be reformulated as a Chern-Simons theory in which the
diffeomorphisms are transmuted into ordinary gauge transformations,
and the results from Chern-Simons theory apply directly.

In particular, for Lorentzian gravity with a negative cosmological
constant $\Lambda=-1/\ell^2$, we can define an $\hbox{SU}(1,1)\times
\hbox{SU}(1,1)$ gauge field
\beq
A^\pm = \left(\tilde\omega^a \pm{1\over\ell}\tilde e^a\right)\tilde T_a ,
\label{b8l}
\eeq
where $\tilde e^a = \tilde e^a{}_\mu dx^\mu$ is a triad and $\tilde\omega^a
= {1\over2}\epsilon^{abc}\tilde\omega_{\mu bc}dx^\mu$ is a spin connection.
The standard Einstein action can then be written as
\beq
I_{\hbox{\scriptsize grav}}
  = I_{\hbox{\scriptsize CS}}[A^+] - I_{\hbox{\scriptsize CS}}[A^-] ,
\label{b9l}
\eeq
where $I_{\hbox{\scriptsize CS}}[A]$ is the Chern-Simons action \rref{b1}
with a coupling constant\footnote{I take $\tilde T_0=i\sigma_3/2$,
$\tilde T_1=\sigma_1/2$, $\tilde T_2=\sigma_2/2$, with $\hbox{Tr}$
the matrix trace.  This normalization differs from that of \cite{Car:stat}.
In the notation of \cite{Witten2}, my $k$ is $is$, with $s$ pure imaginary.
Equation \rref{b10} can be checked by comparing the extremal action for a
closed manifold in the metric and Chern-Simons formalisms.}
\beq
k = -{\ell\over4G} .
\label{b10}
\eeq
We can now continue to Euclidean signature by setting $e^3=i\tilde e^0$,
$e^1=\tilde e^1$, $e^2=\tilde e^2$.  The action then becomes
\beq
I_{\hbox{\scriptsize grav}}
  = I_{\hbox{\scriptsize CS}}[A] - I_{\hbox{\scriptsize CS}}[\bar A] ,
\label{b9}
\eeq
where
\beq
A = \left(\omega^a + {i\over\ell} e^a\right) T_a , \qquad
\bar A = \left(\omega^a - {i\over\ell} e^a\right) T_a
\label{b8}
\eeq
is an $\hbox{SL}(2,{\bf C})$ gauge field (with $T_a=-i\sigma_a/2$).
Our goal is to count the boundary states in this theory for the
three-dimensional black hole.

\section{Quantization}

$\hbox{SL}(2,{\bf C})$ is a noncompact group, and the techniques developed
for quantizing Chern-Simons theories with compact gauge groups require
some modification.  However, the action \rref{b9} looks tantalizingly
like the difference between two $\hbox{SU}(2)$ Chern-Simons actions.
It is thus tempting to treat $A$ and $\bar A$, and the corresponding
gauge transformations $g$ and $\bar g$, as independent fields, and
write
\beq
Z_{\hbox{\scriptsize SL}(2,{\bf C})}[\tilde A, \skew5\bar{\tilde A}]
  = \left| Z_{\hbox{\scriptsize SU}(2)}[\tilde A] \right|^2 ,
\label{c1}
\eeq
where $Z$ denotes the partition function for the WZW action \rref{b4}
on $\partial M$.  Note that since the integrand in the partition
function is $\exp\{iI\}$, the complex conjugation in \rref{c1}
automatically leads to the difference in sign between the two terms
in the action \rref{b9}.

Witten has shown that this procedure is essentially correct \cite{Witten2}.
If one chooses a real polarization, which in our case amounts to fixing
$A_z$ and $\bar A_{\bar z}$ at $\partial M$, then the dependence of a
wave function on $A$ and $\bar A$ is determined entirely by its dependence
on $\omega$.  In particular, it is sufficient to evaluate the partition
function at $e^a=0$ and then ``analytically continue.''  But for $e^a=0$,
the two terms in \rref{b9} are ordinary $\hbox{SU}(2)$ Chern-Simons actions,
and wave functions are basically products of two conjugate $\hbox{SU}(2)$
wave functions.  Hayashi has worked out the resulting $\hbox{SL}(2,{\bf C})$
wave functions for the solid torus in great detail \cite{Hayashi}, and has
shown explicitly that a basis can be constructed from products of holomorphic
affine $\hbox{SU}(2)$ Weyl-Kac characters (from the first $\hbox{SU}(2)$)
and their complex conjugates (from the second).\footnote{In references
\cite{Witten2,Hayashi} the focus was on Chern-Simons states, but a simple
reinterpretation extends the results to the partition function.  Viewed as
a functional of boundary data, a Chern-Simons state on $\partial M$ may be 
defined as a path integral over $M$, weighted by appropriate Wilson lines;
the partition function on a manifold with boundary is thus formally
equivalent to a particular state.}  

Note, however, that in the standard quantization of an $\hbox{SU}(2)$ WZW
model, and in Hayashi's computations, the coupling constant $k$ must be a
positive integer---that is, by \rref{b10}, we must analytically continue to
negative $G$.  This sign change is identical to that described by Henningson
et al.\ \cite{Henningson}, who show that the partition function for an
$\hbox{SU}(1,1)$ WZW model is formally identical to an $\hbox{SU}(2)$ WZW
partition function analytically continued to $k\!<\!-2$.  To obtain a final
answer for the entropy in the Lorentzian theory, we will therefore start
with the Euclidean partition function \rref{c1} with positive integral $k$,
and continue to negative $k$ at the end of the computation.

The advantage of the Euclidean approach is that the path integral for an
$\hbox{SU}(2)$ WZW theory is well understood.  In particular, if
$\partial M$ is a two-torus with modulus $\tau=\tau_1+i\tau_2$, the
partition function $Z_{\hbox{\scriptsize SU}(2)}[\tilde A]$ can be
described as follows \cite{EMSS,LaBas,Falceto}.  We first perform a
gauge transformation to set the gauge field $\tilde A_z$ on $\partial M$
to a constant value
\beq
a = - {\pi i\over\tau_2}u\, T_3
\label{c1a}
\eeq
in the Cartan algebra.  Then for $k$ a positive integer,
\beq
Z_{\hbox{\scriptsize SU}(2)}[\tilde A]
  = \sum_{n=0}^{k} \psi_{nk}(0)\psi_{nk}(a)
\label{c2}
\eeq
with
\beq
\psi_{nk}(a) = \exp\left\{ {\pi k\over4\tau_2}\,\bar u^2 \right\}
  \bar\chi_{nk}(\bar\tau, \bar u) ,
\label{c3}
\eeq
where $\chi_{nk}$ are the Weyl-Kac characters for affine $\hbox{SU}(2)$.
Later we will need the asymptotic behavior of the characters for large
$\tau_2$:
\beq
\chi_{nk}(\tau,u) \sim \exp\left\{
  {\pi i\over2}\left[{(n+1)^2\over k+2} - {1\over2}\right]\tau \right\}
  {\sin\pi(n+1)u\over\sin\pi u} .
\label{c4}
\eeq

Our interest is not the partition function per se, but the number of
states.  For the partition function on a torus with modulus $\tau$,
standard WZW theory \cite{GepWit} tells us that
\beq
Z_{\hbox{\scriptsize SL}(2,{\bf C})}(\tau)[\tilde A, \skew5\bar{\tilde A}]
  = \hbox{Tr}\left\{ e^{2\pi i\tau L_0}e^{-2\pi i\bar\tau \bar L_0}\right\}
  = \sum \rho(N,\bar N)q_1^{N-\bar N}q_2^{N+\bar N} ,
\label{c5}
\eeq
where $q_1=e^{2\pi i\tau_1}$, $q_2=e^{-2\pi\tau_2}$, and $\rho(N,\bar N)$
is the number of states for which the Virasoro generators $L_0$ and $\bar
L_0$ have eigenvalues $N$ and $\bar N$.  This number can be extracted from
\rref{c5} by a standard contour integral:
\beq
\rho(N,\bar N) = -{1\over 4\pi^2}\int{dq_1\over q_1^{N-\bar N+1}}
  \int{dq_2\over q_2^{N+\bar N +1}}
  Z_{\hbox{\scriptsize SL}(2,{\bf C})}(\tau)[\tilde A, \skew5\bar{\tilde A}] ,
\label{c6}
\eeq
where the integrals are along circles surrounding the origin in the
complex $q_1$ and $q_2$ planes.

\section{The Euclidean Black Hole}

We are now ready to count the states of the three-dimensional Euclidean
black hole.  Note first that not all states on the black hole horizon
are physical.  As we saw above, the diffeomorphisms generated by Killing
vectors---vectors in the kernel of $K$---remain genuine gauge symmetries
even at a boundary.  For the BTZ black hole, Killing vectors generate
time translations and rotations, and the corresponding requirement on
states is that
\beq
L_0|\hbox{phys}\!>\, = \bar L_0|\hbox{phys}\!>\, = 0 ,
\label{d0}
\eeq
since the Virasoro operators $L_0$ and $\bar L_0$ generate the rigid
displacements.  Equation \rref{d0} can be viewed as a remnant of the
Wheeler-DeWitt equation.  The number of states at the horizon is thus
given by $\rho(0,0)$.

We next need the boundary fields $\tilde A$ and $\skew5\bar{\tilde A}$.
For this purpose, the black hole metric is most conveniently expressed
in an upper half-space form \cite{CarTeit},
\beq
ds^2 = {\ell^2\over R^2\sin^2\chi}
  \left[dR^2 + R^2d\chi^2 + R^2\cos^2\chi d\theta^2 \right] ,
\label{d1}
\eeq
with the identifications
\beq
(\ln R,\theta,\chi)
  \sim (\ln R, \theta+\Theta,\chi)
  \sim (\ln R+{2\pi r_+\over\ell}, \theta+{2\pi|r_-|\over\ell},\chi) .
\label{d2}
\eeq
Here $r_\pm$ are the Euclidean continuations of the radii of the outer
and inner horizons, and $2\pi-\Theta$ is the deficit angle of the
conical singularity at the (Euclidean) horizon; the on-shell condition
is $\Theta=2\pi$.  The relationship of the coordinates $(R,\theta,\chi)$
to standard Schwarzschild coordinates is described in \cite{CarTeit}.
For our purposes, we need only know that $\chi$ is related to the usual
radial coordinate, and that a surface $\chi=\hbox{const.}$ is a torus
(the two circumferences are a circle around the horizon and a circle in
periodic time); the horizon is the degenerate surface $\chi=\pi/2$.

The connection $A^a$ corresponding to the metric \rref{d1} is easily
found to be
\beq
A^1 = -\csc\chi (d\theta - i{dR\over R}) ,\qquad
A^2 = i\csc\chi d\chi ,\qquad
A^3 = i\cot\chi (d\theta - i{dR\over R}) .
\label{d3}
\eeq
When restricted to a ``stretched horizon'' $\chi=\chi_0$, $A$ is conjugate
to $(d\theta - i{dR\over R})T_3$, independent of $\chi_0$.  To use this
boundary data in the partition function, we must express this connection
as $a(dx+\tau dy)$, where $x$ and $y$ are coordinates on the torus with
period one.  Using the identifications \rref{d2}, or equivalently rewriting
$a$ in terms of the holonomies of $\tilde A$, we obtain
\beq
a =
  -{\pi\over\tau_2}\left[\left({\Theta\tau_2\over2\pi}+{r_+\over\ell}\right)
  +i\left({\Theta\tau_1\over2\pi}-{|r_-|\over\ell}\right) \right] T_3 .
\label{d4}
\eeq

For $k$ a positive integer, we can now insert this expression, along with
the asymptotic form \rref{c4} of the Weyl-Kac characters, into \rref{c3}
to compute the partition function $Z_{\hbox{\scriptsize SL}(2,{\bf C})}$.
The integral \rref{c6} may then be evaluated by steepest descent.  Consider,
for example, the contribution from $n=0$ in \rref{c2}.  The corresponding
term in the partition function is
\begin{eqnarray}
Z_{\hbox{\scriptsize SL}(2,{\bf C})}[a,\bar a]
&\approx& \exp\left\{ {\pi k\over4\tau_2}\,(u^2 + \bar u^2)
  + {\pi k\over k+2}\tau_2 \right\} \nonumber\\
&=& \exp\left\{ -{\pi k\over2\tau_2}\left[
  \left( {\Theta\tau_2\over2\pi}+{r_+\over\ell}\right)^2
  - \left( {\Theta\tau_1\over2\pi}-{|r_-|\over\ell}\right)^2\right]
  + {\pi k\over k+2}\tau_2 \right\} .
\label{d5}
\end{eqnarray}
The steepest descent approximation of integral \rref{c6} for $N=\bar N=0$
then gives
\beq
\rho(0,0) = \exp\left\{ -{k\Theta r_+\over\ell}
  + \left({2\pi\over\Theta}\right){\pi r_+\over\ell} \right\}
  = \exp\left\{ {\Theta r_+\over4G}
  + \left({2\pi\over\Theta}\right){\pi r_+\over\ell} \right\}
\label{d6}
\eeq
up to terms of order $1/k$.  Note that the relevant saddle point occurs
at $\tau_2=2\pi r_+/\Theta\ell$, so the approximation \rref{c4} is
justified as long as the black hole is large, $M/G = 8r_+^2/\ell^2\gg1$.
A straightforward calculation also shows that the contributions to
$\rho(0,0)$ coming from terms in \rref{c2} with $n\ne0$ are exponentially
suppressed relative to \rref{d6}.

(There is a sign ambiguity here: in the identifications \rref{d2}, we
could have taken $r_+$ and $|r_-|$ to be negative, corresponding
to a different fundamental region.  The relevant saddle point would
then be $\tau_2=-2\pi r_+/\Theta\ell$, and the first term in \rref{d6}
would no longer appear.  For this choice, however, it may be shown that
the contribution coming from $n=k$ in \rref{c2} reproduces \rref{d6},
with $\Theta$ replaced by $4\pi-\Theta$.)

The first term in the exponent of \rref{d6} is the correct semiclassical
expression for the entropy of the (2+1)-dimensional black hole.  The
second term is a one-loop correction.  This one-loop expression differs
from that of reference \cite{CarTeit} by a factor of two, but I believe it
is correct; the expression in \cite{CarTeit} was based on a computation of
determinants in reference \cite{Car:entropy} (eqn.\ A16) which, I believe,
has an incorrect factor of two in the exponent.

The Euclidean computation has been carried out for $k$ a positive integer,
implying that $G\!<\!0$.  As noted above, however, the analytic continuation
to Lorentzian signature requires a change of the sign of $k$.  It might
be possible to repeat this computation directly in the Lorentzian theory,
using the results of reference \cite{Henningson} for the $\hbox{SU}(1,1)$
partition function.  In fact, though, the semiclassical contribution to
the entropy is largely independent of the detailed form of the characters
$\chi_{nk}(\tau,u)$.  The leading contribution to $S$ comes from the prefactor
\beq
Z_0(\tau,u)=\exp\left\{ {\pi k\over4\tau_2}\,(u^2 + \bar u^2)\right\}
\label{d7}
\eeq
in the partition function \rref{d5}.  This term may be understood as
follows.  In a Chern-Simons theory on a manifold with boundary, the
WZW current $J(z)$ at the boundary is proportional to the gauge field
$A_z$ \cite{Banados}.  By \rref{b2}, this field contains the usual
current $g^{-1}\partial g$, but it also has an added zero-mode
contribution $\tilde A_z$, whose value is determined by the boundary
data.  The Virasoro generator $L_0$ has a corresponding zero-mode term
proportional to $\hbox{Tr}\tilde A_z\tilde A_z$, and the prefactor
\rref{d7} is precisely the contribution of this zero-mode to the
partition function.  As originally argued in reference \cite{Car:stat},
it is this zero-mode that determines the dependence of the black hole
entropy on the horizon size.

This means that although the entropy counts microscopic states at
the horizon, its semiclassical value will not depend sensitively on
the quantum theory describing those states, as long as the current
zero-modes are fixed.  For example, although we do not know the
detailed form of the characters corresponding to $\chi_{nk}(\tau,u)$
for $k$ nonintegral, and we do not fully understand the Lorentzian
theory, the semiclassical expression for black hole entropy should
not be affected: provided the partition function has the form
$Z_0(\tau,u)F(\tau,u)$ with $F\sim1$ for large $\tau_2$, we will obtain
the correct Bekenstein-Hawking entropy.\footnote{From \rref{c5} and its
generalizations, this condition on $F$ should hold as long as the number
of WZW states does not increase too rapidly with $N$ and $\bar N$.}

Finally, let me return to the question of whether it is sensible to
treat a black hole horizon as a boundary.  The horizon is not, of course,
a physical boundary.  It is, however, a place at which one must impose
``boundary conditions'' in quantum gravity.  Statements about black
holes in quantum gravity are necessarily statements about conditional
probabilities: for instance, ``If a black hole with given characteristics
is present, then one will observe a certain spectrum of Hawking radiation.''
To compute such probabilities, one must include the appropriate restrictions
on the path integral, by restricting the admissible boundary data at the
horizon.  Such restrictions are sufficient to generate a WZW action at the
horizon, and thus to justify the computations of this paper \cite{PCGC}.

\section{Conclusion}

We have seen that the Chern-Simons formulation of three-dimensional
Euclidean gravity permits an explicit description of horizon degrees
of freedom, and that these degrees of freedom can provide a microscopic
explanation for the entropy of the black hole.  The obvious question
is whether these results can be generalized to four dimensions.  The
particular methods described here certainly cannot.  The key advantage
of the Chern-Simons formalism is that it allows diffeomorphisms to be
expressed as local gauge transformations, permitting the decomposition
\rref{b2} and the exact derivation of a boundary action.  No such
formulation is known in 3+1 dimensions.

Nevertheless, the basic physical mechanism discussed here should
generalize to 3+1 dimensions.  The canonical formulation of general
relativity offers strong evidence for the existence of ``would-be gauge''
degrees of freedom that can become dynamical at a boundary \cite{Bal},
and some progress has been made towards finding the corresponding
boundary action \cite{Teit,Epp}.  While much work remains, the approach
developed here provides a promising direction for understanding the
origin of black hole entropy.

\vspace{1.5ex}
\begin{flushleft}
\large\bf Acknowledgements
\end{flushleft}

This work was supported in part by National Science Foundation grant
PHY-93-57203 and Department of Energy grant DE-FG03-91ER40674.

\end{document}